# BEFANA: A Tool for Biodiversity-Ecosystem Functioning Assessment by Network Analysis

Martin Marzidovšek[1, 3*], Vid Podpečan[1], Erminia Conti[2], MarkoDebeljak[1,3], Christian Mulder[2]

1 Jozef Stefan Institute, Department of Knowledge Technologies, University of Ljubljana (Slovenia)

2 Department of Biological, Geological and Environmental Sciences, University of Catania (Italy)

3 Jožef Stefan International Postgraduate School (Slovenia)

## Summary

1. BEFANA is a free and open-source software tool for ecological network analysis and visualisation. It is adapted to ecologists' needs and allows them to study the topology and dynamics of ecological networks as well as apply selected machine learning algorithms.

2. BEFANA is implemented in Python, and structured as an ordered collection of interactive computational notebooks. It relies on widely used open-source libraries, and aims to achieve simplicity, interactivity, and extensibility.

3. BEFANA provides methods and implementations for data loading and preprocessing, network analysis and interactive visualisation, modelling with experimental data, and predictive modelling with machine learning. We showcase BEFANA through a concrete example of a detrital soil food web of agricultural grasslands, and demonstrate all of its main components and functionalities.

---

[*] Correspondence author. E-mail: martin.marzidovsek@ijs.si. Present address: Jozef Stefan Institute, Jamova 39, 1000 Ljubljana, Slovenia.

## Introduction

Unsustainable use of natural resources have been compelling the scientific community to turn its interest to biodiversity loss in general, and to negative cascading effects in particular. Several works concerning ecological networks have been published[1] but too few of them are mechanistic and predictive, despite the plea by Ings et al. (2009). On the other hand, online dataset collections and international research interest is rapidly growing, especially for soil organisms (van den Hoogen et al. 2020; Phillips et al. 2021).

Data on the ecological interactions in food webs are generally represented in four different ways: I) Resource-to-consumer food web, where the structural model is either static or dynamic, with directed graphs (flow) for functional groups, emphasising energy channels, loops, biomass and nutrient cycling (Hunt and Wall 2002; Fitter 2005; Mulder et al. 2005); II) Allometric scaling (size spectra), where the resulting power law is often environmentally-driven, showing triangular plots for taxa, trophic generality and vulnerability, trophic heights, assimilation efficiency, addressing stoichiometric imbalances (Reuman and Cohen 2004; Mulder and Elser 2009); III) Topological networks, with a static biopyramid for species, trophospecies and/or functional groups, emphasising structural biodiversity, size-delimited pathways, nested hierarchy, and/or parasitism (Cohen, Jonsson, and Carpenter 2003; 2009; Hudson et al. 2013; Mulder et al. 2005) and finally, IV) Interaction networks for species, trophospecies and/or functional groups, focusing on connectance, complexity, trophic transfer efficiency, self-damping, distribution of trophic links and chain lengths, magnitude of feeding relationships etc. (Reuman and Cohen 2004; Kondoh, Kato, and Sakato 2010; Bascompte and Jordano 2007).

---

[1] 443 papers recorded on Web of Science as of February 15, 2022, using the specific query: AK=(ecological network*) AND AB=(ecological network*) AND TI=(ecological network*) NOT AK=(social).

This work presents BEFANA, a free and open-source software tool suitable for ecological network analysis and visualisation for resource-to-consumer food webs, and topological and interaction networks (the aforementioned architectural types I, III and IV). It allows the ecologist to study the topology of the network and its temporal dynamics, identify indirect interactions and more. Ease of use, interactivity, and extensibility were the main guidelines when designing and developing BEFANA. As a result, BEFANA is implemented as a collection of interactive computational notebooks and relies on widely used open-source libraries for network analysis, data science and machine learning. BEFANA is available through a public web interface as well as a Docker container and local installation[2]. The rest of the paper showcases selected parts of BEFANA through a case study of detrital soil food-web analysis.

## Design and implementation

The design of BEFANA (Fig. 1) follows the latest advances in developing and sharing of scientific software. It is based on the computational notebook paradigm and follows the "batteries included" philosophy by providing integration of libraries a user might need in ecological network analysis. As such, BEFANA is an interactive software tool but also a solid base for creating custom applications and experiments in this and related domains.

BEFANA is implemented in the Python programming language, which has become a *de facto* language of choice for scientific computing (Perez & Granger 2007). In combination with JupyterLab, an interactive development environment for computational notebooks (i.e. Jupyter notebooks), and pandas (McKinney 2010), a Python library for data manipulation and analysis offering optimised data structures and functions, it forms one of the best approaches to the dissemination of scientific programs, experiments, data manipulation and interactive computational notebooks. BEFANA is based on an open-source software stack which helps to achieve the following three goals: (a) ease of use, (b) interactivity, and (c) extensibility.

---

[2] MyBinder: https://mybinder.org/v2/gh/MartinMarzi/BEFANA/v1.2.1
GitHub: https://github.com/MartinMarzi/BEFANA

The libraries currently included in BEFANA cover the following topics: data input and output, preprocessing and manipulation, charting, web-based graphical user interfaces, network analysis and visualisation, and machine learning. It is an open collection as any other libraries can be applied in BEFANA notebooks as long as they implement an interface to the Python interpreter or are available as a web service or a command line tool

Even though it is common for computational notebooks to be run locally, BEFANA goes a step further and also provides a public web interface through *mybinder.org*, a free public service which provides a community-led infrastructure deploying the BinderHub technology (Bussonnier et al. 2018), which allows sharing of interactive code. Local installation of BEFANA is possible using Docker or by creating a virtual Python environment where BEFANA can be started.

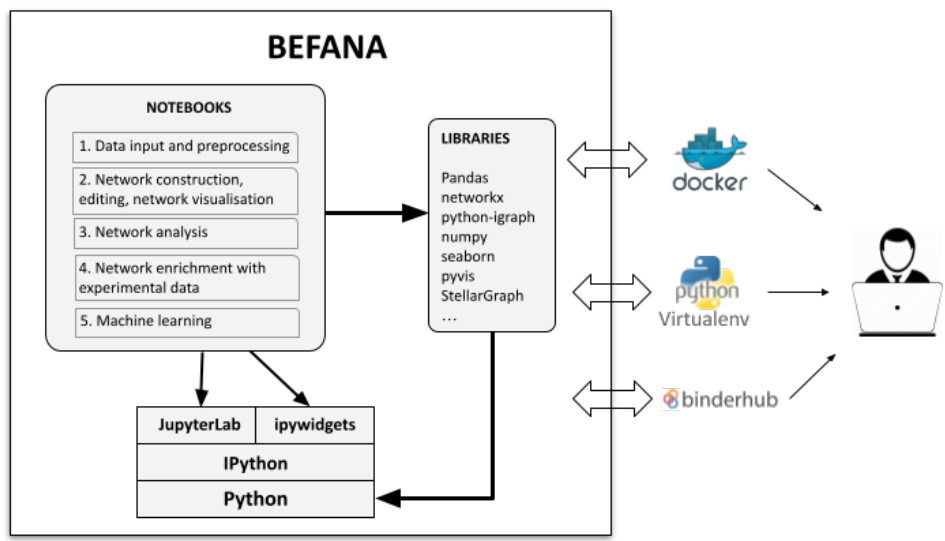

Figure 1: Architecture of BEFANA.

## Tool demonstration through soil food-web analysis

In BEFANA's computational notebooks, code, interface and results, including diagrams and graphics interweave and offer a uniquely rich user experience. BEFANA can be used to demonstrate some particular implementation, to showcase an example, to share charts with results or to provide network visualisations. In this section we demonstrate the application of BEFANA on a case of a soil food-web data where we quantify and analyse the hierarchical characteristics of the trophic network at hand.

A hierarchical directed graph has been constructed, representing the detrital soil food web using one regional metaweb of all possible literature-based multitrophic links from a local pool of food webs (Mulder et al. 2005; Sechi et al. 2015). The complete analysis consists of the following: (1) data input and preprocessing, (2) network construction, editing and network visualisation, (3) network analysis, (4) network enrichment with experimental data, and (5) machine learning.

**Launching BEFANA**

The easiest and recommended way to work with BEFANA is to use the public Binder link[3]. Local deployment using Docker or a Python virtual environment is recommended for advanced users. Moreover, BEFANA includes already computed results of the presented soil food web use case for each notebook. All BEFANA notebooks ensure that the required libraries are installed prior to executing any code.

---

[3] MyBinder: https://mybinder.org/v2/gh/MartinMarzi/BEFANA/v1.2.1

**Data input and preprocessing**

BEFANA supports several different input formats including csv, Excel, Access and HDF. In our example, the input data are loaded as a csv file with headers and a custom separator. If the input data requires preprocessing steps such as imputation, filtering, resampling, computing values etc. this is the right place to apply them.

**Network construction and editing**

The data from which a network is constructed can be provided in several different ways. For example: (a) the network can be loaded from a file in one of supported network description languages (e.g. Pajek, GML, JSON), (b) it can be built in the Python code by adding nodes and links manually, (c) it can be constructed from a data structure such as the adjacency matrix or adjacency list. In our example the network is constructed from the adjacency matrix of the metaweb, encompassing three local plots, hereafter called A, B, and C (Sechi et al. 2015).

**Network visualisation and exploration**

One of the strongest features of BEFANA is its interactive visualisation of hierarchical networks. In ecosystem network analysis visualisation is an essential tool for understanding the structure of the network, the interrelations between nodes of interest and for hypothesis formation. BEFANA implements interactive visualisation of ecological networks and responds to user events such as zoom and drag. A network can be visualised either inline in the notebook or in a new browser tab. Inline visualisation follows the computational notebook paradigm and displays a preview of the network and few customization options in a notebook cell while the advanced visualisation in a new tab offers a multitude of options such as colours, shapes, sizes, fonts, layouts and their optimization parameters etc.

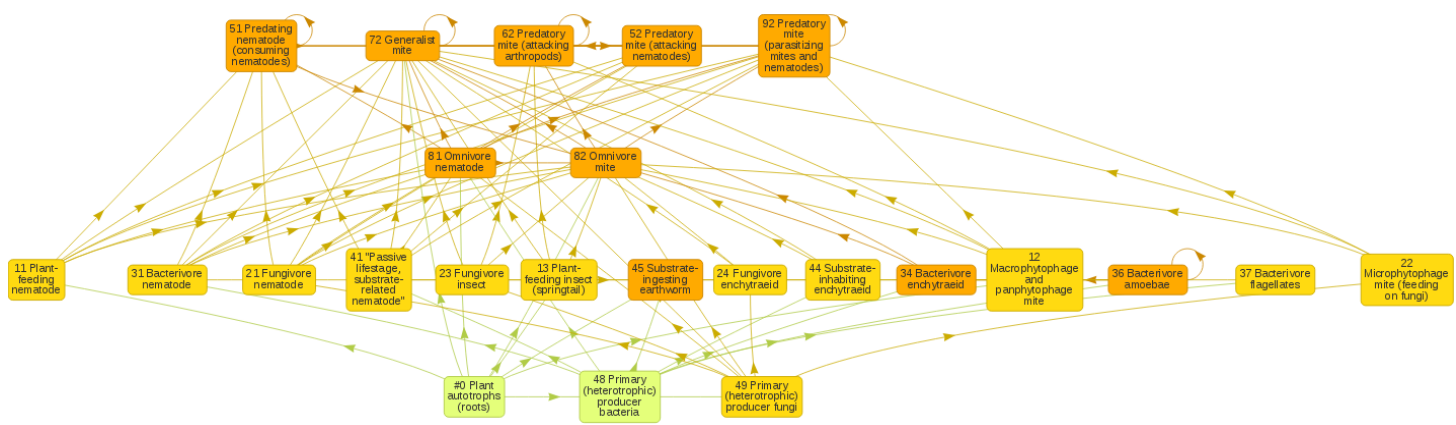

Figure 2: A hierarchical visualisation of the soil metaweb with a directed graph in BEFANA. Trophic levels define the layers of the network and the colouring of nodes provides a visualisation of their functional roles. Nodes represent trophic groups and their links to the resources: for example, plant roots are present at the bottom and provide food for plant-feeding nematodes, generalist mites, omnivore nematodes, and a number of other trophic groups. In this way, omnivores and top predators that consume from more than one node and/or more than one trophic level are easily recognizable. Colouring can be done either at functional level (e.g. according to an in-degree clustering rule as presented in the Network analysis section) or at taxonomic level (for instance, different colours at the Kingdom level). The numbers prefixing node names are functional codes for trophic nodes: the first digit provides information on the dominant feeding strategy while the second digit provides cladistic information.

BEFANA's visualisation also enables modelling the relationship between network topology and network dynamics, which has implications in understanding stability and resilience of trophic networks (Dale 2017). It's graphical user interface offers network editing, so users can easily perform network modification and analysis cycles. They can remove certain nodes or edges of a soil food web and then visualise the modified network to assess the extent to which the investigated ecosystem will become destabilised (Murall, McCann, and Bauch 2012), for instance by functional biodiversity decline (Conti et al. 2020). This loss of nodes will impact the subsequent calculations and will enable us to forecast various scenarios of biodiversity loss.

**Network analysis**

Once the network construction and visualisation are complete, a user can proceed with the network analysis that allows to characterise the form of the ecological network and determine how the functioning of the network affects its structure (Dale 2017). While species interactions are central, graph theory can also reveal higher level (guild to guild) interactions, such as how topology of such networks also interacts with the functioning of the ecological system.

NETWORK CHARACTERISTICS

Keystone trophic groups and connections that have a disproportionate effect on the network function are identified through the combination of various centrality measures that give insight on the nodes/edges topological position and role within the network in a similar way to Cheddar (Hudson et al. 2013). Figure 5 shows a selection of centrality measures. Each measure provides different insights on trophic node's properties or position in the ecological network.

In addition, the out/in-degree ratio (van den Brink and Rusinowska 2021) has been calculated, which assigns to every node in a directed graph its out-degree divided by its in-degree (to avoid dividing by zero, 1 is added to both the out- as well as in-degree of every node). It can be used for ranking nodes in a directed graph. This measure is similar to the well-known Copeland score for ranked voting but can give different results in certain situations. Interestingly, in our soil food web structure one node from the second trophic level (bacterivore nematodes) is ranked the same as one of the primary resources (fungi), and higher than other second level nodes.

COMPARTMENT AND MOTIF ANALYSIS

The soil food web was analysed for common repeated structural elements and how clearly separable such compartments are. Pairs of nodes with reciprocal link were identified, which also includes cannibalistic relationships. For example, our soil food web contains 9 reciprocal trophic interactions (interspecific loops) of which 5 are cannibalistic (intraspecific loops), mostly occurring at higher trophic levels.

All transitive triangle motifs were also discovered, and an example is visualised in Figure 3. Transitivity depends on directional interactions and is related to the consistency of relationships among nodes (Dale 2017). Thus, transitive triangles hint on the stability of the trophic network as they demonstrate the presence of more stable combinations of trophic interactions where a trophic group has more than one resource. Our soil metaweb contains 122 triangles describing transitive relationships.

Strongly connected components can give an indication about the energy flow within the network. Interestingly, our soil food web contains exactly one strongly-connected component spanning three nodes, which is shown in Figure 3. Moreover, this component (embedding only groups from trophic level 4) also contains 8 of 9 possible links which implies a very high resistance to biodiversity loss. As such, it contains a cycle spanning all three nodes, two cycles spanning two nodes (reciprocal links) and three loops (all three trophic groups are cannibalists). One of the trophic groups is a generalist consumer while the other two are predators.

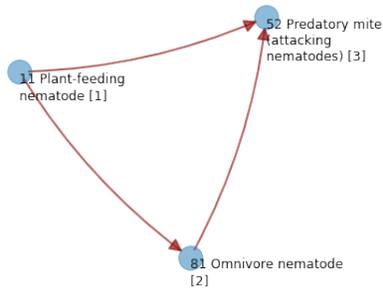

Figure 3a: An example of a transitive triangle spanning three trophic levels (the number in square brackets denotes trophic level).

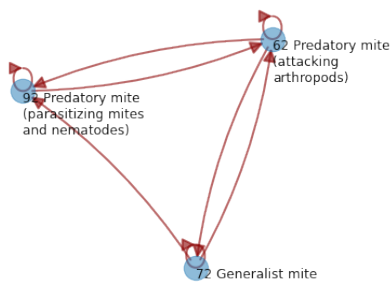

Figure 3b: The most strongly-connected component in our soil food web.

Finally, in order to discover meaningful subgraphs, clustering within the network can be performed. The results of clustering (node partitioning) can also be used in BEFANA's network visualisation. For example, in-degree based node clustering into three clusters where indegree is either 0, 1, or greater than 1 yields clusters as shown in Figure 2. Such clustering separates trophic groups with energy sources from outside of the network, highly vulnerable trophic groups having only one source of energy and less vulnerable having more than one source of energy. With the exception of substrate ingesting earthworms, bacterivore enchytraeid, and bacterivore amoebae all trophic groups on the first level are vulnerable, in contrast to trophic groups on the second and third trophic level (the average indegree is 8.2).

*BEFANA: A Tool for Ecological Network Analysis*DERIVED NETWORKS

An important advantage of using BEFANA on ecological networks is its ability to identify the existence and strengths of indirect interactions that are derived from the existing trophic network. BEFANA's computational notebook #4 implements the discovery and visualisation of two types of derived networks: (1) with shared consumer, and (2) with shared resource. In the analysed soil food web (1) was very connected with 83% of nodes linking to over 60% of all the nodes. In (2), on the other hand, centrality trends in the derived soil food network were clearer: only three nodes had a degree centrality over 60% (omnivore nematode, generalist mite, and omnivore mite).

**Modelling with experimental data**

BEFANA allows the user to embed experimental or field data on the network structure which can provide insight on the dynamics of the studied ecosystem. The aggregated biomass, mass and abundance per trophic group can be computed and added to the network as node attributes. This allows us to easily compute the total biomass for all trophic levels. Any additional data preprocessing, specific to the problem and data at hand, can be also performed at the same time. The network data structure allows for an unlimited number of attributes to be assigned to nodes and links. Attributes can also be used in computations provided by the NetworkX library (where applicable) simply by providing the name of the relevant attribute. Moreover, experimental data can also be used in training machine learning models.

The abundances, average body-size values and biomass values of trophic groups on all trophic levels can be easily computed and ranked. According their biomass, showing that earthworms and the three main resources for the plant-driven pathway, the bacterial-driven pathway and the fungal-driven pathway depict the highest biomass values in each of the three plots of the investigated metaweb, whilst the other groups are ranked differently according to the plot, and two are not even occurring in all plots (Figure 5). It is worth mentioning that all groups in common in all separate food webs share a very low coefficient of variation (CV) (less than 25%), in contrast to the group ranked as 21st, that although occupying the same rank according to its biomass values shares a CV higher than 100%. BEFANA also implements quantitative comparison of ranked lists of biomass values using the rank-biased overlap (RBO) measure (Webber, Moffat, and Zobel 2010) which measures the similarity of potentially infinite ranked lists. It has one parameter (p) which determines how top-weighted the measure is: the smaller p, the more it is top-weighted. Applying RBO to the ranked list of biomass values of all three plots yields the following results. Regardless of the value of p, plot A is always the most similar to plot C. The least similar are plots A and B if RBO is top-weighted (low p), or plots B and C, if the tails of the biomass value lists are also considered in the comparison (high p). The RBO-based similarity of all three plots using different values of p is shown in Figure 4.

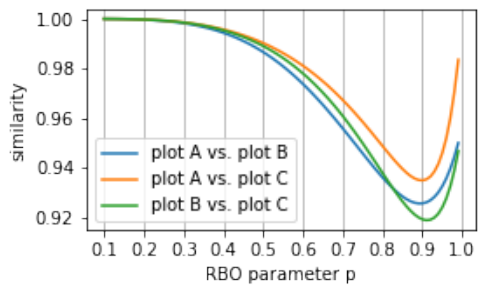

Figure 4: Similarity of ranked lists of biomass values for local plots A, B, and C using the RBO measure and different values of its top-weightedness parameter *p*.

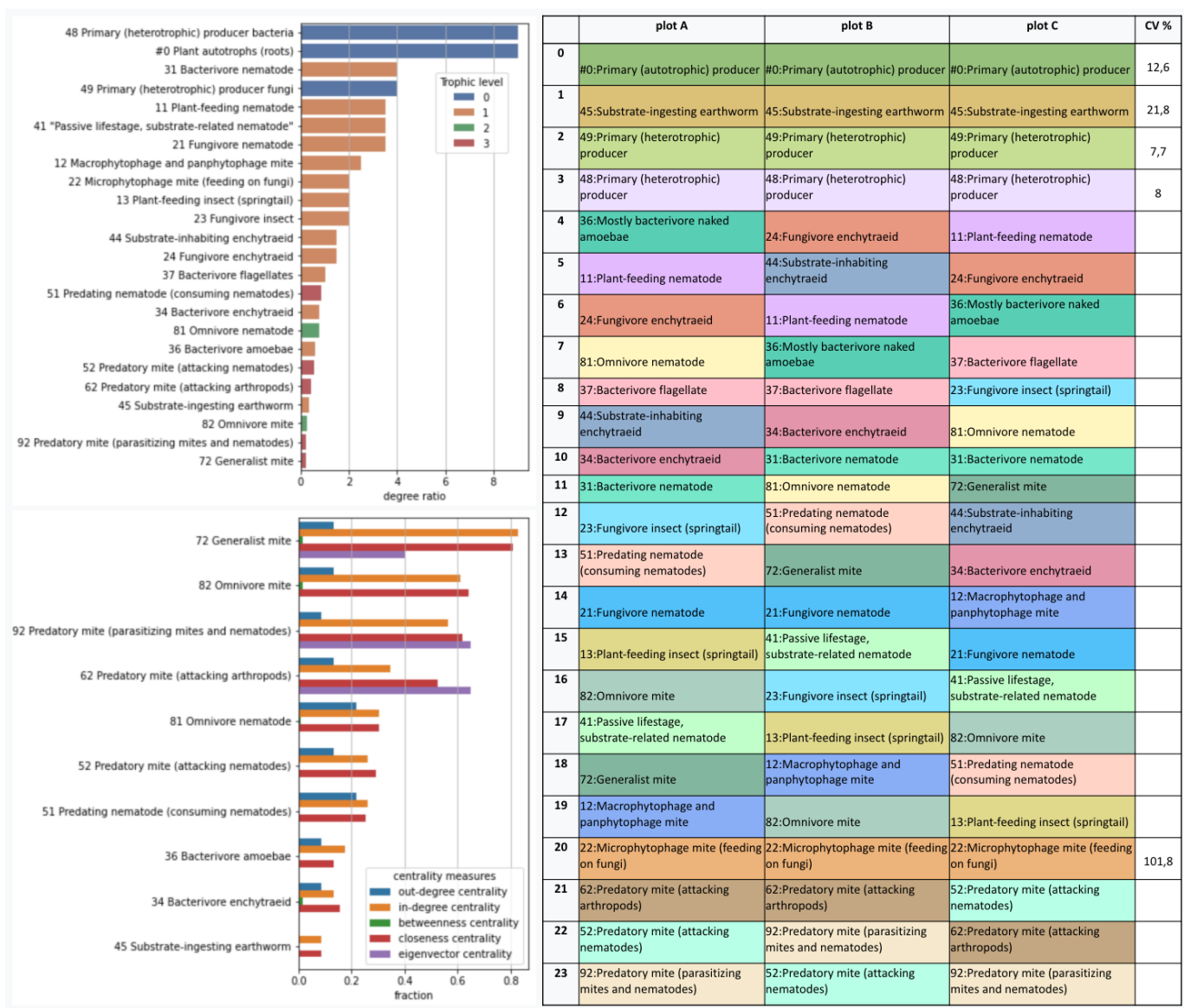

Figure 5. The top-left figure shows a selection of centrality measures of top-10 nodes. The bottom-left figure shows the degree ratio according to in-degree centrality. On the right the ranked biomass averages are shown for each separate local plot (A, B and C). The Coefficient of Variation has been computed only for the biomass averages of any trophic group ranked at the same position across the metaweb, five trophic groups in our case. Each colour (plant: green, fungi: brown, bacteria: blue) resembles the pathway, with yellow and reddish for higher trophic level groups (omnivores and predators).

**Representation learning**

To employ network data in predictive and descriptive machine learning tasks (e.g. classification, link prediction, clustering, graph evolution) the data (nodes, edges) has to be represented as numerical vectors or node embeddings. Node2vec is one of the recent machine learning algorithms that learns such a mapping of nodes to a low-dimensional space of features while maximising the likelihood of preserving network neighbourhoods of nodes (Grover & Leskovec 2016). BEFANA implements this representation learning method and applies it to the derived network described above. The results demonstrate that the topological information is indeed encoded in node embeddings, and dimensionality reduction followed by the visualisation in 1D space gives a clearer indication of node similarity. As can be seen on Figure 6, one of the primary producers (fungi) is further apart and thus different from the other two (bacteria and plant autotrophs) as its energy source is within the network. The four trophic groups on the far left in Figure 6 that are clearly separated from the rest are the only nodes at trophic level 1 sharing fungi as resource.

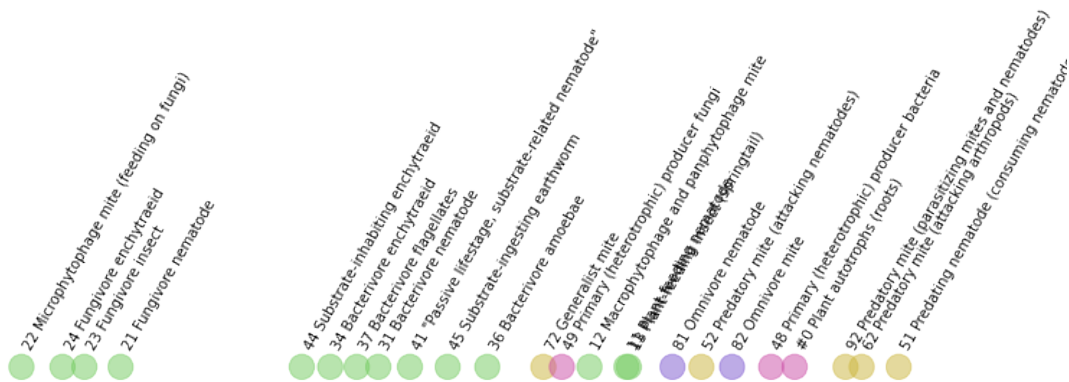

Figure 6. Visualisation of node embeddings in 1D where the distance between the trophic groups is indicative about node similarity in the derived network. Colours represent the trophic level (e.g. red is for primary producers).

## Conclusion and future development

The paper presented BEFANA, a free and open-source software tool for ecological network visualisation and analysis. It allows ecologists to take advantage of graph theory and machine learning on ecological networks that can help quantifying and testing hypotheses about how a network's topology affects its dynamics and how the functioning of the network in turn affects its structure. In the future work, we plan to upgrade BEFANA regularly with new features, implementations and datasets. A particular focus will be on machine learning methods optimised for ecological networks. Experimental data is important in ecological analysis and to make use of them machine learning algorithms need to incorporate node and/or edge attributes. While the hierarchical structure of our soil food web makes it unsuitable for representation learning methods, we envision that node embeddings for other types of interaction networks could well be used to predict links when simulating ecological network disruptions.

## Acknowledgements

This work has been supported by COST action 18237 and project H2020 project EMBEDDIA (#825153). We thank Prof. Dr. Nada Lavrač for her kind support.

## Conflict of interest

Nothing to declare.

## Data availability

Dryad data: http://dx.doi.org/10.5061/dryad.t5347.

## Author contributions

MM and VP designed the software; CM and EC contributed with data and workflow improvements; CM, MD and EC verified the outcomes; MM, VP, MD and CM led the writing of the manuscript; all authors contributed critically and improved for publication.

## References

Bascompte, J., & Jordano, P. (2007). Plant-Animal Mutualistic Networks: The Architecture of Biodiversity. *Annual Review of Ecology, Evolution, and Systematics* 38 (1): 567–593.

Brink, R. van den, & Rusinowska, A. (2021). The Degree Ratio Ranking Method for Directed Graphs. *European Journal of Operational Research* 288 (2): 563–575.

Cohen, J.E., Jonsson, T. & Carpenter, S.R. (2003). Ecological Community Description Using the Food Web, Species Abundance, and Body Size. *Proceedings of the National Academy of Sciences* 100(4): 1781–1786.

Cohen, J.E., Schittler, D.N., Raffaelli, D.G., & Reuman, D.C. (2009). Food Webs Are More than the Sum of Their Tritrophic Parts. *Proceedings of the National Academy of Sciences* 106(52): 22335–22340.

Conti, E., Di Mauro, L.S., Pluchino, A., & Mulder, C. (2020). Testing for Top-down Cascading Effects in a Biomass-Driven Ecological Network of Soil Invertebrates. *Ecology and Evolution* 10(14): 7062–7072.

Dale, M.R.T. (2017). *Applying Graph Theory in Ecological Research*. Cambridge University Press.

Fitter, A.H. (2005). Darkness Visible: Reflections on Underground Ecology. *Journal of Ecology* 93(2): 231–243.

Grover, A., & Leskovec, J. (2016). node2vec: Scalable Feature Learning for Networks. In *Proceedings of the 22nd ACM SIGKDD International Conference on Knowledge Discovery* *and Data Mining*, 855–64. KDD'16. New York, NY.

Hagberg, A., Swart, P., & Chult, D.S. (2008). Exploring Network Structure, Dynamics, and Function Using NetworkX. Los Alamos National Laboratory, Los Alamos, NM. https://www.osti.gov/biblio/960616.

Hoogen, J. van den, Geisen, S., et al. (2020). A Global Database of Soil Nematode Abundance and Functional Group Composition. *Scientific Data* 7(1): 103.

Hudson, L.N., et al. (2013). Cheddar: Analysis and Visualisation of Ecological Communities in R. *Methods in Ecology and Evolution* 4(1): 99–104.

Hunt, H.W., & Wall, D.H. (2002). Modelling the Effects of Loss of Soil Biodiversity on Ecosystem Function. *Global Change Biology* 8(1): 33–50.

Ings, T.C., et al. 2009. Ecological Networks--beyond Food Webs. *Journal of Animal Ecology* 78 (1): 253–69.

Kondoh, M., Kato, S., & Sakato, Y. (2010). Food Webs Are Built up with Nested Subwebs. *Ecology* 91 (11): 3123–3130.

McKinney, W. (2010). Data Structures for Statistical Computing in Python. *Proceedings of the 9th Python in Science Conference.* DOI:10.25080/Majora-92bf1922-00a

Mulder, C., & Elser, J.J.. (2009). Soil Acidity, Ecological Stoichiometry and Allometric Scaling in Grassland Food Webs. *Global Change Biology* 15(11): 2730–2738.

Mulder, C., Cohen, J.E., Setälä, H., Bloem, J., & Breure, A.M. (2005). Bacterial Traits, Organism Mass, and Numerical Abundance in the Detrital Soil Food Web of Dutch Agricultural Grasslands. *Ecology Letters* 8(1): 80–90.

Murall, C.L., McCann, K.S., & Bauch, C.T. (2012). Food Webs in the Human Body: Linking Ecological Theory to Viral Dynamics. *PLoS One* 7(11): e48812.

Perez, F., & Granger, B.E. (2007). IPython: A System for Interactive Scientific Computing. *Computing in Science & Engineering.* DOI:10.1109/mcse.2007.53.

Phillips, H.R.P., et al. (2021). Global Data on Earthworm Abundance, Biomass, Diversity

and Corresponding Environmental Properties. *Scientific Data* 8(1): 136.

Project Jupyter et al. (2018). Binder 2.0 - Reproducible, Interactive, Sharable Environments for Science at Scale. *Proceedings of the 17th Python in Science Conference*. DOI:10.25080/majora-4af1f417-011.

Reuman, D.C., & Cohen, J.E. (2004). Trophic Links' Length and Slope in the Tuesday Lake Food Web with Species' Body Mass and Numerical Abundance. *Journal of Animal Ecology* 73(5): 852–866.

Sechi, V., Brussaard, L., De Goede, R.G.M., Rutgers, M., & Mulder, C. (2015). Choice of Resolution by Functional Trait or Taxonomy Affects Allometric Scaling in Soil Food Webs. *American Naturalist* 185(1): 142–149.

Webber, W., Moffat, A., & Zobel, J. (2010). A Similarity Measure for Indefinite Rankings. *ACM Transactions on Information and System Security* 28(20): 1–38.